\documentclass[twoside]{article}
\usepackage{fleqn,espcrc2}

\def\beq{\begin{equation}}
\def\eeq{\end{equation}}
\def\bea{\begin{eqnarray}}
\def\eea{\end{eqnarray}}
\def\bq{\begin{quote}}
\def\eq{\end{quote}}

\def\fr{\frac}
\def\ve{\varepsilon}
\def\dt{\Delta t}
\def\dm{\Delta m}
\def\dg{\Delta \Gamma}
\def\nn{\nonumber}
\newcommand{\ol}{${\cal O}(\lambda^3)$}

\newcommand{\AmS}{{\protect\the\textfont2
  A\kern-.1667em\lower.5ex\hbox{M}\kern-.125emS}}

\title{Indirect Violation of CP, T and CPT in the $B_d$-system}

\author{M.C. Ba\~nuls\address{IFIC (Centro Mixto Univ. Valencia - CSIC)
 	46100 Burjassot (Valencia), Spain}
 }      
\begin{document}

\begin{abstract}
The problem of indirect violation of discrete symmeties CP, T and CPT in a neutral meson system can be described using two complex parameters $\varepsilon$ and $\delta$, which are invariant under rephasing of meson and quark fields.
For the $B_d$ system, where the width difference between the physical 
states is negligible, only ${\rm Re}(\delta)$ and 
${\rm Im}(\varepsilon)$ survive. 
As a consequence, the traditional observables constructed for kaons, which are based on flavour tag, are not useful for the analogous study in this system.
We describe how using a CP tag and
studying \emph{CP-to-flavour} transitions of the $B$ mesons, we may build 
asymmetries, alternative to those used for the kaon, which 
enable us to test T and CPT invariances of the effective hamiltonian 
for the $B_d$ system.
\vspace{1pc}
\end{abstract}

\begin{titlepage}
\onecolumn
\begin{flushright} {FTUV-00-0927}
\end{flushright}
\vskip 4cm
\centerline{\LARGE \bf
Indirect Violation of CP, T and CPT in the $B_d$-system
\footnote{To appear in the Proceedings of 4th International Conference on Hyperons, Charm and Beauty Hadrons, Valencia (Spain) 27-30 June 2000.}
}
\vskip 1cm
\centerline{
M.C. Ba\~nuls}
\vskip 0.5cm
\centerline{
{IFIC, Centro Mixto Univ. Valencia - CSIC}}

\vskip 1cm
\centerline{\large \bf Abstract}
\vskip 0.5 cm
The problem of indirect violation of discrete symmeties CP, T and CPT in a neutral meson system can be described using two complex parameters $\varepsilon$ and $\delta$, which are invariant under rephasing of meson and quark fields.
For the $B_d$ system, where the width difference between the physical 
states is negligible, only ${\rm Re}(\delta)$ and 
${\rm Im}(\varepsilon)$ survive. 
As a consequence, the traditional observables constructed for kaons, which are based on flavour tag, are not useful for the analogous study in this system.
We describe how using a CP tag and
studying \emph{CP-to-flavour} transitions of the $B$ mesons, we may build 
asymmetries, alternative to those used for the kaon, which 
enable us to test T and CPT invariances of the effective hamiltonian 
for the $B_d$ system.

\end{titlepage}
\newpage

\maketitle

\section{Introduction}
The time evolution of a neutral meson system is governed by
an effective hamiltonian~\cite{ka68}.
The problem of indirect violation of discrete symmetries 
refers to the non-invariance of this hamiltonian under 
the corresponding operations.

For the kaon system, this study has been performed by the CP-LEAR 
experiment~\cite{cplear} from the preparation of 
definite flavour states $K^0$-$\bar{K}^0$.
These tagged mesons evolve in time and their later decay to
a semileptonic final state projects them again on a definite 
flavour state.
The study of this \emph{flavour-to-flavour} evolution 
allows the construction of
observables which violate CP and T, or CP and CPT.

Contrary to what happens in the kaon case, for the $B_d$-system
the width difference $\dg$ between the physical states is expected
to be negligible.
In this system the T- and CPT-odd observables proposed for kaons,
which are based on flavour tag, vanish.
but, making use of CP tag, the $B_d$ entangled states can be used to
construct alternative observables which are sensitive to
T and CPT independently of the value of $\dg$~\cite{bb99.2}.

\section{The parameters}

In the neutral $B$-meson system the physical states are a
linear combination of $B^0$ and $\bar{B}^0$.
If they are written in terms of CP eigenstates, one has to introduce
two complex parameters, $\ve_{1,2}$, to describe the CP mixing.
\beq 
|B_{\stackrel{\scriptstyle 1}{\scriptstyle 2}} \rangle = 
\frac{1}{\sqrt{1+|\varepsilon_{\stackrel{\scriptstyle 1}{\scriptstyle 2}}|^2}} \left [|B_{\pm}
\rangle +
\varepsilon_{\stackrel{\scriptstyle 1}{\scriptstyle 2}}
|B_{\mp} \rangle \right ] \ ,
\label{fis}
\eeq
where $|B_{\pm}\rangle \equiv \frac{1}{\sqrt{2}}(I \pm CP)|B^0\rangle$.
Then $\ve_{1, 2}$ are invariant under rephasing of the meson states, 
and physical when the CP operator is well defined~\cite{bb98}.

Alternatively, one may use the parameters $\ve \equiv (\ve_1+\ve_2)/2$ 
and $\delta \equiv \ve_1-\ve_2$, whose interpretation in terms of 
symmetries is simpler.

Discrete symmetries impose different restrictions
on the effective mass matrix, $H=M-\frac{i}{2}\Gamma$:
CPT invariance requires $H_{11}=H_{22}$, 
T invariance imposes 
${\rm Im}(M_{12} {\rm CP}_{12}^*)={\rm Im}(\Gamma_{12}{\rm CP}_{12}^*)=0$,
and CP conservation requires both conditions to be simultaneously satisfied.
Furthermore, in the exact limit $\Delta \Gamma=0$, customary for the 
$B_d$-system, both ${\rm Re}(\ve)$ and ${\rm Im}(\delta)$ vanish.
Therefore we have four real parameters which carry information on 
the symmetries of the effective mass matrix
\begin{itemize}
\item{${\rm Re}(\varepsilon) \Rightarrow$ CP and T violation, with 
$\dg\!\neq\!0$;}
\item{${\rm Im}(\ve) \Rightarrow$ CP and T violation;}
\item{${\rm Re}(\delta) \Rightarrow$ CP and CPT violation;}
\item{${\rm Im}(\delta) \Rightarrow$ CP and CPT violation, 
$\dg \!\neq \!0$.}
\end{itemize}

\section{The entangled state: CP tag}

In a $B$ factory operating at the $\Upsilon(4S)$ peak, 
correlated pairs of neutral $B$-mesons are produced through
$e^+ e^- \rightarrow \Upsilon(4S) \rightarrow B \bar{B}$.
The special features of this system can be used to 
study CP~\cite{wo84} and CPT~\cite{ko92} violation 
in $B$ mesons.

In the CM frame, the resulting $B$-mesons travel in opposite directions, 
each one evolving with the effective hamiltonian.
The $B \bar{B}$ state has definite $L=1$, $C=-$ and ${\cal P}=-$,
being $\cal P$ the operator which permutes the spatial coordinates,
so that the initial state may be written as
\beq
\vert i>=\frac{1}{\sqrt{2}} 
\left (\vert B^0, \overline{B}^0>
- \vert \overline{B}^0, B^0> \right ) 
\label{eq:ent}
\eeq

The correlation between both sides of the entangled 
state holds at any time after the production.
As a consequence, one can never simultaneously have two 
identical mesons at both sides of the detector.
This permits the performance of a flavour tag: if at $t=0$ 
one of the mesons decays through a channel, such as a semileptonic one,  
which is only allowed for one flavour of the neutral $B$,
the other meson in the pair must have the opposite flavour at $t=0$.

The entangled $B-\bar{B}$ state can also be expressed 
in terms of the CP eigenstates 
$|B_{\pm}\rangle$ as
\beq
\vert i>=\frac{1}{\sqrt{2}} \left (\vert B_-, B_+>
- \vert B_+, B_-> \right )
\label{eq:entCP}
\eeq
Thus it is also possible to carry out a CP tag, 
once we have a CP-conserving decay into a definite CP final state, 
so that its detection allows us to identify the decaying meson 
as a $B_+$ or a $B_-$.

In Ref.~\cite{bb99} we described how this determination is possible
and unambiguous to \ol,
which is sufficient to discuss both CP-conserving and CP-violating
amplitudes in the effective hamiltonian for $B_d$ mesons.
Here $\lambda$ is the flavour-mixing parameter of the CKM matrix~\cite{ko73}.
The determination is based on the requirement of CP
conservation, to \ol, in the $(sd)$ and $(bs)$ sectors.
To this order, however, CP-violation exists in the $(bd)$ sector, 
and it can be classified by referring it to the CP-conserving direction.
A $B_d$ decay that is governed by the couplings of the
$(sd)$ or $(bs)$ unitarity triangles, or by the $V_{cd} V_{cb}^*$ side
of the $(bd)$ triangle, will not show any CP violation to \ol.
We may say that such a channel is free from direct CP violation.
Examples are $J/\Psi K_S$, with ${\rm CP}=-$, and $J/\Psi K_L$, 
with ${\rm CP}=+$.

To extract information on the symmetry parameters we may
study the time evolution of the entangled state (\ref{eq:ent})
and its decay into a final configuration $(X,\, Y)$.
In our notation, $X$ is the decay product observed 
on one side of the detector at a certain time, and $Y$ the product detected 
on the opposite side after a $\dt$.

We will only consider here decay channels $X$, $Y$ which are either 
flavour or CP conserving.
Then the final configuration $(X,\,Y)$ corresponds to a certain transition 
at the mesonic level, i.e. the $B$ state tagged by the $X$ decay
evolves for a period $\Delta t$ and is then projected into a 
flavour or CP eigenstate by means of the $Y$ decay.

\section{The asymmetries}

By comparing the probabilities corresponding to different processes 
we build time-dependent asymmetries that allow 
the extraction of the relevant parameters. 
The observables can be classified into three types.

\subsection{Flavour-to-flavour genuine asymmetries}

If one detects semileptonic decays on both sides of the detector,
then the transition at the meson level is of the kind 
\emph{flavour-to-flavour}.
The mesonic transitions for such a final
configuration appear in Table~\ref{tab:f2f}, where $\ell^{\pm}$ 
represents the final decay product of a semiinclusive decay 
$B \rightarrow \ell^{\pm} X^{\mp}$.
\begin{table}[htb]
\caption{
\emph{Flavour-to-flavour} transitions}
\label{tab:f2f}
\begin{tabular}{cc}
\hline
$(X,\, Y)$ & Transition \\
\hline 
$(\ell^+, \, \ell^+)$ & $\bar{B}^0 \stackrel{\phantom{c}}{\rightarrow} {B^0}$ \\
$(\ell^-, \, \ell^-)$ & $B^0 \rightarrow \bar{B}^0$ \\
$(\ell^+, \, \ell^-)$ & $\bar{B}^0 \rightarrow \bar{B}^0$ \\
$(\ell^-, \, \ell^+)$ & $B^0 \rightarrow B^0$ \\
\hline
\end{tabular}
\end{table}
From these processes we can construct two non-trivial asymmetries,
which are the analogous, in the $B$-system, to the traditional 
observables used for kaons.
The first two processes in Table~\ref{tab:f2f}
are conjugated under CP and also under T, then we may construct a 
genuine asymmetry by comparing the corresponding intensities
\beq
A (\ell^+, \, \ell^+) \approx 
{\scriptstyle \fr{{\rm Re}(\ve)}{1+ |\ve|^2}}.
\label{eq:l+l+}
\eeq

On the other hand, the last two processes in Table~\ref{tab:f2f}
are related by a CP or a CPT transformation.
Therefore, the corresponding asymmetry, 
\bea
A(\ell^+ , \, \ell^-) & \!\!\approx \!\!&\!\!
- 2\left[ {\rm Ch} {\scriptstyle\frac{\dg \dt}{2}}\!+\! \cos({\scriptstyle\Delta m \dt})\right]^{-1} \nn \\
&& \hspace{-1.8cm}
\left[{\rm Re} \!\left (\! {\scriptstyle\frac{\delta}{1-\varepsilon^2}}\!\right ) {\rm Sh} {\scriptstyle\frac{\Delta \Gamma \Delta t}{2}} 
\!-\! {\rm Im} \!{\scriptstyle\left ( \!\frac{\delta}{1-\varepsilon^2}\!\right )} \sin({\scriptstyle \Delta m \Delta t})\right ],
\label{eq:l+l-}
\eea
is also a genuine CP and CPT observable.

In both cases, the resulting asymmetry vanishes unless $\dg \neq 0$.
Thus measuring a small value for these observables does not 
impose a straightforward bound on
the size of symmetry violation, because the 
vanishingly small $\dg$ of $B$-mesons would hide any symmetry 
breaking effect.

\subsection{CP-to-flavour genuine asymmetries}

We may construct alternative asymmetries 
making use of the CP eigenstates, 
which can be identified in this system by means of a CP tag.
If the first decay product, $X$, is a CP eigenstate
produced along the CP-conserving direction, and $Y$ is
a semileptonic channel, then the mesonic transition corresponding
to the configuration $(X, \, Y)$ is of the type \emph{CP-to-flavour}.
The order of appearance of both final states
matters, because for the reverted configuration, $(Y,\, X)$,
we have a \emph{flavour-to-CP} transition.
In Table~\ref{tab:CP2f} we show the mesonic transitions, with their 
related final configurations, connected by genuine symmetry transformations
to $B_+ \rightarrow B^0$, i.e. $(J/\Psi K_S, \, \ell^+)$.
\begin{table}[htb]
\caption{
Transitions connected to $(J/\Psi K_S, \, \ell^+)$.}
\label{tab:CP2f}
\begin{tabular}{ccc}
\hline
$(X,\, Y)$ & Transition & Transformation \\ 
\hline
$(J/\Psi K_S, \, \ell^-)$ & $B_+ \rightarrow \bar{B}^0$ & CP\\
$(\ell^-, \, J/\Psi K_L)$ & $B^0 \rightarrow B_+$ & T\\
$(\ell^+, \, J/\Psi K_L)$ & $\bar{B}^0 \rightarrow B_+$ & CPT\\
\hline
\end{tabular}
\end{table}
Comparing the intensity of $(J/\Psi K_S, \, \ell^+)$ with each 
of them we construct three genuine asymmetries.
Next, we show the results to linear order in $\delta$ 
and in the limit $\dg=0$.
\bea
A_{{\rm CP}} & \! = \!& 
-2 {\scriptstyle \fr{{\rm Im}(\ve)}{1+|\ve|^2}} \sin ({\scriptstyle \dm \dt}) 
\nn \\
&&  \hspace{-.5cm}
+ {\scriptstyle \fr{1-|\ve|^2}{1+|\ve|^2}} 
{\scriptstyle \fr{2 {\rm Re}(\delta)}{1+|\ve|^2}} 
\sin^2 \left ({\scriptstyle \fr{\dm\dt}{2}}\right ),
\label{eq:aCP}
\eea
is the CP odd asymmetry,
which has contributions from T-violating and CPT-violating terms.
The first term, odd in $\dt$, is governed by the T-violating ${\rm Im}(\ve)$,
whereas the second term, $\dt$ even, is sensitive 
to CPT violation through the parameter ${\rm Re}(\delta)$.
\bea
A_{\rm T} &=& 
-2 {\scriptstyle \fr{{\rm Im}(\ve)}{1+|\ve|^2}} \sin ({\scriptstyle\dm \dt}) 
\nn \\
&& \hspace{-.5cm}
 \left[ 1 - {\scriptstyle \fr{1-|\ve|^2}{1+|\ve|^2}} 
{\scriptstyle \fr{2 {\rm Re}(\delta)}{1+|\ve|^2}} 
\sin^2 \left ({\scriptstyle \fr{\dm \dt}{2}}\right) \right],
\label{eq:aT}
\eea
the T asymmetry, 
needs $\ve \neq 0$, and includes CPT even and odd terms.
Moreover, in the limit we are considering, turns out to be 
purely odd in $\dt$. 
\beq
A_{{\rm CPT}}=
{\scriptstyle \fr{1-|\ve|^2}{1+|\ve|^2}} 
{\scriptstyle \fr{2 {\rm Re}(\delta)}{1+|\ve|^2}} 
\fr{\sin^2 \left ({\scriptstyle\fr{\dm \dt}{2}}\right )}
{1-2 \fr{{\rm Im}(\ve)}{1+|\ve|^2} \sin ({\scriptstyle\dm \dt})},
\label{eq:aCPT}
\eeq
is the CPT asymmetry. It
needs $\delta \neq 0$, and includes both even and odd time dependences,
so that there is no definite symmetry under a change of sign of $\dt$.

Measuring the presented asymmetries (\ref{eq:aCP})-(\ref{eq:aCPT})
with good time resolution, so to separate even and odd $\dt$ dependences,
should be enough to determine the parameters
$\fr{2 {\rm Im}(\ve)}{1+|\ve|^2}$
and
$\fr{1-|\ve|^2}{1+|\ve|^2} \fr{2 {\rm Re}(\delta)}{1+|\ve|^2}$,
which govern CP, T violation and CP, CPT violation, respectively, 
in the $B_d$ mixing.

Contrary to what happened in the case of flavour tag, the
CPT and T asymmetries based on a CP tag do not vanish due
to the smallness of $\dg$.
Instead, they provide a set of observables which could separate
the parameters $\delta$ and $\ve$.

\subsection{CP-to-flavour non-genuine asymmetries}

The asymmetries defined in the previous paragraphs are genuine
observables, since each of them compares the original process with
its conjugated under a certain symmetry and is thus odd under 
the corresponding transformation.
Nevertheless the measurement of all those quantities requires
to tag both $B_+$ and $B_-$ states.
The last needs, from the experimental point of view, a good reconstruction 
of the decay $B \rightarrow J/\Psi K_L$, not so easy to achieve as
for the corresponding $J/\Psi K_S$ channel.

But it is also possible to construct useful asymmetries 
from final configurations $(X, \, Y)$ with only $J/\Psi K_S$.
\begin{table}[htb]
\caption{
Final configurations with only $J/\Psi K_S$.}
\label{tab:CP2fng}
\begin{tabular}{ccc}
\hline
$(X,\, Y)$ & Transition & Transformation \\ 
\hline
$(J/\Psi K_S, \, \ell^-)$ & $B_+ \rightarrow B^0$ & CP\\
$(\ell^+, \, J/\Psi K_S)$ & $\bar{B}^0 \rightarrow B_-$ & $\dt$\\
$(\ell^-, \, J/\Psi K_S)$ & $\bar{B}^0 \rightarrow B_-$ & $\dt$+CP\\
\hline
\end{tabular}
\end{table}
In Table ~\ref{tab:CP2fng} we show the different transitions 
we may study from such final states.
From the comparison between $(J/\Psi K_S, \, \ell^+)$ and each 
process in the table we can construct three asymmetries.
The first one will correspond to the genuine CP asymmetry
$A(J/\Psi K_S, \, \ell^-)=A_{\rm CP}$.
We find that, in the exact limit $\dg =0$, $\dt$ and T operations 
become equivalent, so that
$A(\ell^+, \, J/\Psi K_S) \equiv A_{\rm T}$ and
$A(\ell^-, \, J/\Psi K_S) \equiv A_{\rm CPT}$.
But these asymmetries are not genuine.
They do not correspond to true T- and CPT-odd observables, for
the processes we are comparing are not related by a symmetry
transformation. 
This implies that the presence of $\dg \neq 0$ may induce
non-vanishing values for them, 
even if there is no true T or CPT violation.
But even if that is the case, it is possible to separate out the 
different parameters, if good enough $\dt$ is provided ~\cite{bb00}.

\section{Conclusions}

We present an overview of the possibilities to explore indirect 
violation of CP, T and CPT in a neutral meson system from the 
quantities that $B$-factories can measure.
The asymmetries analyzed here exploit their time 
dependences in order to separate out two different ingredients:
on one hand CP and T violation, described by $\ve$, and
on the other CP and CPT violation, given by $\delta$.
Such a study is possible, even if $\dg\!=\!0$, if
one goes beyond \emph{flavour-to-flavour} transitions and
makes use of CP tags.

We classify the observables into three different types:
\begin{itemize}
\item
Genuine asymmetries for T or CPT violation, based on  
{\it flavour-to-flavour} transitions at the meson level,
which need $\dg \! \neq \! 0$.
\item
Genuine observables,
based on the combination of flavour and CP tags, which do not need $\dg$.
\item
Making use of the equivalence between $\dt$ and T reversal operations
for $\dg=0$, 
we have also considered non genuine observables,
involving only the hadronic decay $J/\Psi K_S$. 
\end{itemize}

This work has been supported by CICYT, Spain, under Grant AEN99-0692.
M.C.B. is indebted to the Spanish Ministry 
of Education and Culture for her fellowship.

\end{document}